# LETTER

# Relativistic reverberation in the accretion flow of a tidal disruption event

Erin Kara[1,2,3], Jon M. Miller[4], Chris Reynolds[1,3] & Lixin Dai[3,5]

**Our current understanding of the curved space-time around supermassive black holes is based on actively accreting black holes, which make up only ten per cent or less of the overall population. X-ray observations of that small fraction reveal strong gravitational redshifts that indicate that many of these black holes are rapidly rotating[1]; however, selection biases suggest that these results are not necessarily reflective of the majority of black holes in the Universe[2]. Tidal disruption events, where a star orbiting an otherwise dormant black hole gets tidally shredded and accreted onto the black hole[3], can provide a short, unbiased glimpse at the space-time around the other ninety per cent of black holes. Observations of tidal disruptions have hitherto revealed the formation of an accretion disk and the onset of an accretion-powered jet[4–8], but have failed to reveal emission from the inner accretion flow, which enables the measurement of black hole spin. Here we report observations of reverberation[9–12] arising from gravitationally redshifted iron Kα photons reflected off the inner accretion flow in the tidal disruption event Swift J1644+57. From the reverberation timescale, we estimate the mass of the black hole to be a few million solar masses, suggesting an accretion rate of 100 times the Eddington limit or more[13]. The detection of reverberation from the relativistic depths of this rare super-Eddington event demonstrates that the X-rays do not arise from the relativistically moving regions of a jet, as previously thought[5,14].**

Swift J1644+57 was detected by the Swift Burst Alert Telescope on 28 March 2011 as it reached an isotropic X-ray luminosity greater than $10^{48}$ erg s$^{-1}$. Follow-up radio, optical and infrared observations located the position of the source at the centre of a compact, non-interacting, star-forming galaxy (redshift $z = 0.3534$). The spatial coincidence with a normal galaxy and the decay in luminosity over time $t$ as $L \propto t^{-5/3}$ led to the conclusion that this transient was a tidal disruption event (TDE)[4,6,15]. The peak luminosity exceeds the Eddington limit by four orders of magnitude (for a $10^6$-solar-mass black hole), which suggests that the X-ray emission is highly anisotropic, possibly originating from a relativistic jet pointed in our line of sight[4]. These energetics arguments suggesting a relativistic jet were bolstered by the detection of a strong radio afterglow at the nucleus of this previously inactive galaxy[5]. Although the broadband observations all suggest that this TDE resulted in super-Eddington accretion and the launching of jets, the powering of the jets (whether magnetically driven or radiation-pressure-driven) and the role of geometric beaming remains unknown[14,16].

Follow-up X-ray observations of Swift J1644+57 were taken soon after the initial discovery, and found the source to be highly variable, further suggesting accretion onto a compact object. A 200 s quasi-periodic signal was reported in the 2–10 keV light curve of the brightest two observations at >99.99% and >99.73% confidence[17]. Although the exact nature of the quasiperiodic oscillation is unknown, it does confirm the formation of an accretion disk just 10 days after the initial TDE. It was suggested[17] that this quasiperiodicity could originate from orbits at the inner edge of the accretion disk, resonances in the relativistic jet or variability in a turbulent accretion flow. Here, we reanalyse the early XMM-Newton and Suzaku observations to look at properties of the variability at all other timescales (Extended Data Table 1). We use a new technique, X-ray reverberation mapping, which combines spectral and timing techniques to understand the physical nature of the rapid X-ray variability.

We proceed by extracting XMM-Newton light curves in narrow energy bands from 0.3 keV to 10 keV (observed frame) and then use Fourier techniques to extract the relative time delays between these bands as a function of the temporal frequency (see Methods). We find a coherent soft excess lag in the frequency range 0.0002–0.001 Hz (Extended Data Fig. 1), and by averaging the interband time delays over these frequencies produce the lag–energy spectrum shown in Fig. 1. We find that the 5.5–8 keV rest-frame emission is delayed by ~100 s relative to the 4–5 keV and 8–13 keV emission (the same time delay can be seen directly between the rest-frame 5.5–8 keV and 8–13 keV light curves in Extended Data Fig. 2). The iron K lag is found with >99.9% confidence (see Methods for details). We also find the same iron K lag in the Suzaku observation taken 10 days earlier at >99.8% confidence (Extended Data Figs 3 and 4). The asymmetry of the line is confirmed at >98% confidence (Extended Data Fig. 5, Extended Data Table 2). The iron K lag is characteristic of emission that reverberates off the inner accretion flow and is re-emitted as a relativistically broadened iron K$\alpha$ line. Such short-timescale lags between the direct emission and the reprocessed emission indicate that the emitting region is very small, within tens of gravitational radii of the central black hole. The asymmetric profile of the iron K lag indicates that the iron K$\alpha$ emission line is gravitationally redshifted from within 10 gravitational radii. Relativistic reverberation lags associated with the broad iron K$\alpha$ line are commonly found in rapidly varying local Seyfert galaxies[18–20] (see Extended Data Fig. 6 for examples).

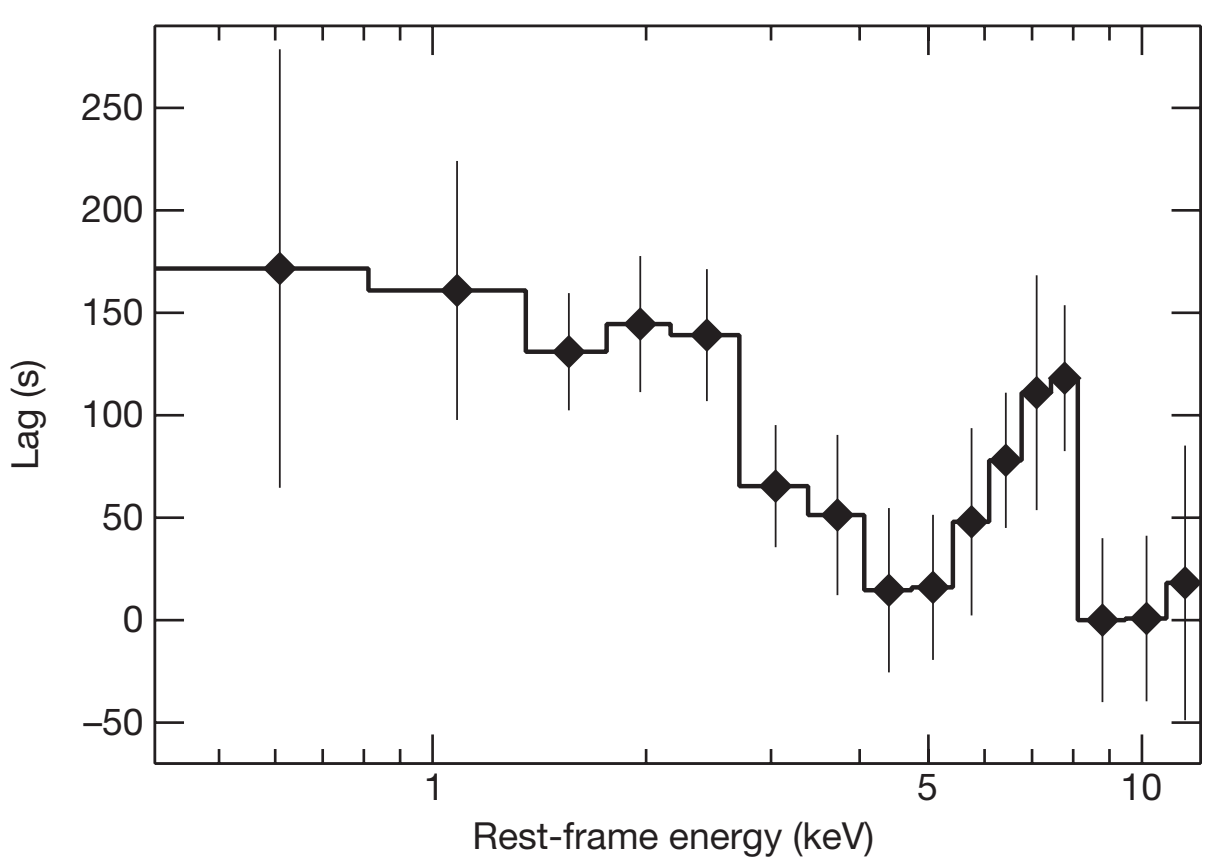

**Figure 1 | Lag–energy spectrum of Swift J1644+57.** The lag–energy spectrum with 1$\sigma$ error bars in the frequency range $(2–10) \times 10^{-4}$ Hz. The zero point lag has been shifted so that zero lag corresponds to the emission that varies first and larger lags correspond to the correlated emission that varies later. This lag–energy spectrum shows that the emission from ~4–5 keV and 8–13 keV (primary continuum-dominated bands) vary first, and the iron line from ~7–8 keV responds ~100 s later, consistent with short-timescale reverberation off an ionized accretion disk. The asymmetric profile of the iron K lag shows that the emission is gravitationally redshifted and therefore much of the emission originates close to the central massive black hole.

[1]Department of Astronomy, University of Maryland, College Park, Maryland 20742, USA. [2]X-ray Astrophysics Laboratory, NASA/Goddard Space Flight Center, Greenbelt, Maryland 20771, USA. [3]Joint Space Science Institute, University of Maryland, College Park, Maryland 20742, USA. [4]Department of Astronomy, University of Michigan, Ann Arbor, Michigan 48103, USA. [5]Department of Physics, University of Maryland, College Park, Maryland 20742, USA.

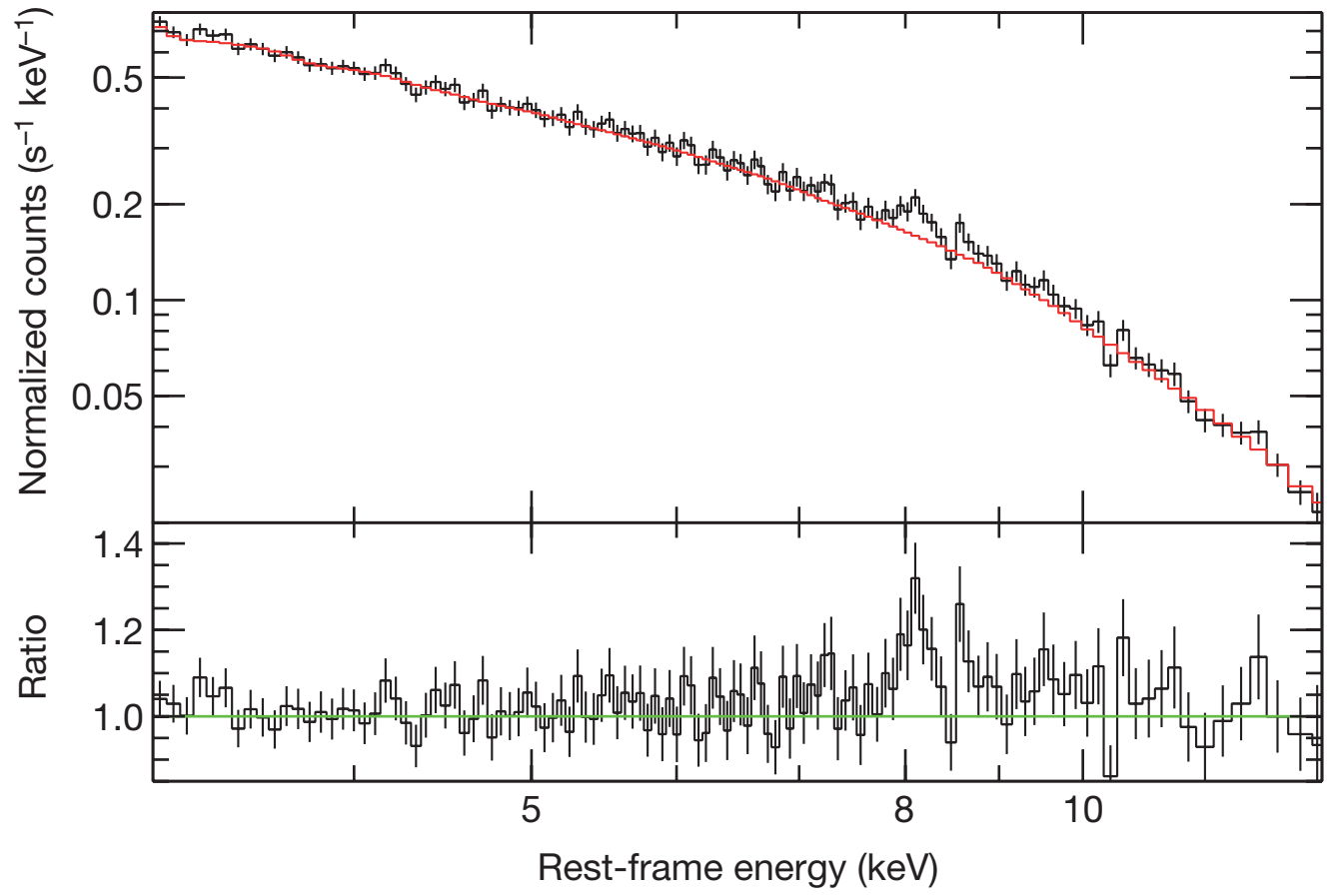

**Figure 2 | The energy spectrum of Swift J1644+57.** The time-integrated energy spectrum in normalized counts units of the observation shown in Fig. 1. The top panel shows the data in black and the best-fit power-law model for 2–10 keV (observer's frame) in red. Errors are at the 1$\sigma$ level. The fit includes intrinsic absorption of $\sim 10^{22}$ cm$^{-2}$, similar to previous analyses[17]. The bottom panel shows the ratio of the data to that of the absorbed power-law model. The green line indicates a ratio of one, corresponding to agreement between the model and the data. The residuals show a clear excess at 8 keV, the same energy as the peak of the lag–energy spectrum in Fig. 1.

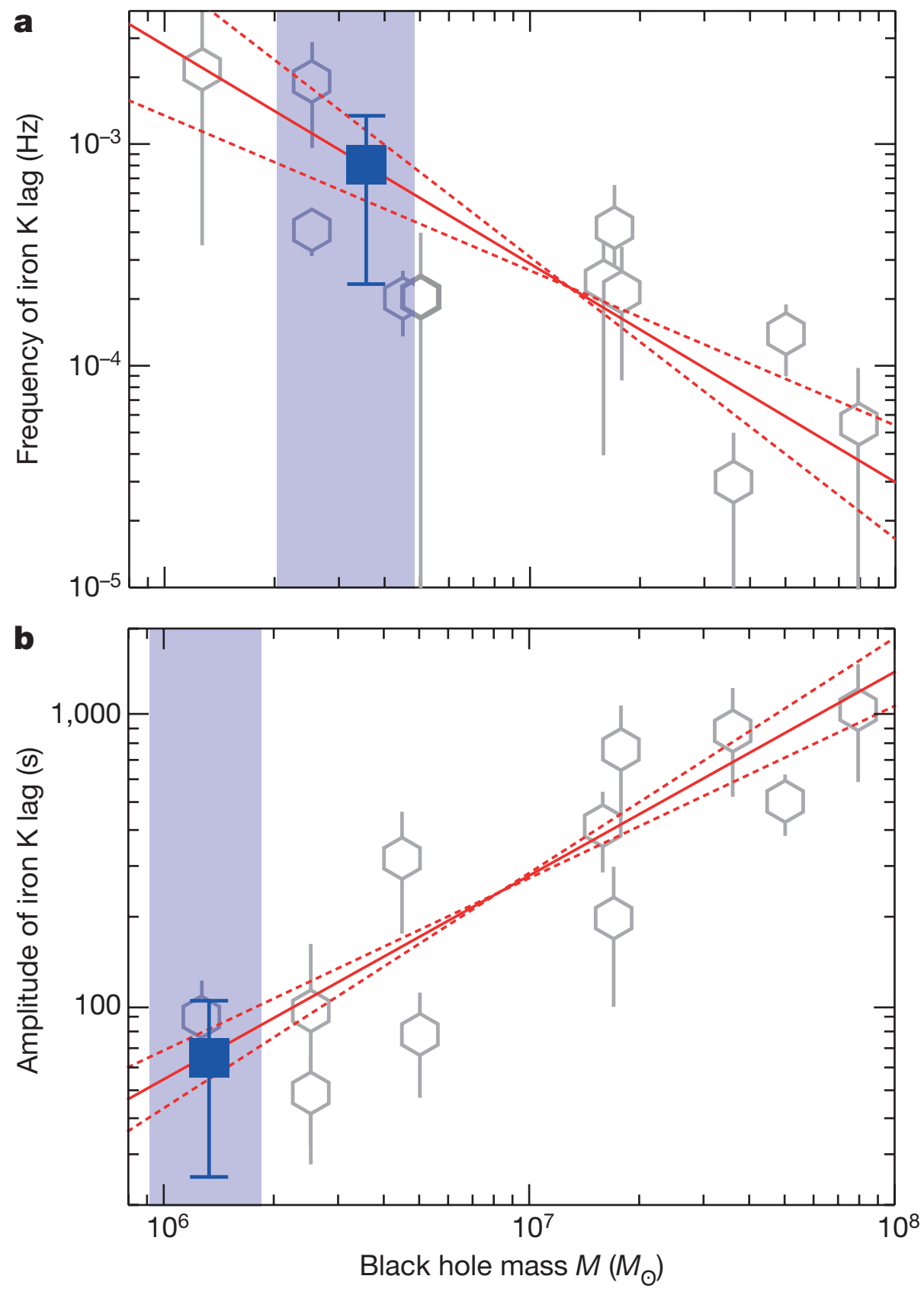

**Figure 3 | Comparison of lags in Swift J1644+57 with other Seyfert galaxies.** **a**, The rest-frame frequency range of the iron K lag versus black hole mass (in units of solar masses, $M_{\odot}$) for several Seyfert galaxies (grey hexagons; error bars indicate the frequency range). The solid red line indicates the best-fit linear model (in log–log space) with 1$\sigma$ errors (red dashed lines). The blue square and error bars indicate the rest-frame frequency range of the iron K lag in Swift J1644+57, and the blue shaded region shows the 1$\sigma$ range in mass suggested by this correlation. **b**, As in **a** except the correlation is with the amplitude of the lag (with 1$\sigma$ error bars) versus black hole mass. The amplitude of the lag in Swift J1644+57 is calculated between 4–5.5 keV and the peak of the line at 6.5–8 keV. Both correlations suggest a black hole mass of a few million solar masses.

Time lag analysis is a powerful technique because it disentangles the variable and correlated emissions from other emissions in the time-integrated energy spectrum. We examined the time-integrated energy spectrum in detail to search for further evidence of the iron K emission line. The resulting spectrum is shown in Fig. 2 compared with the best-fit absorbed power law for 2–10 keV. There is an excess in the energy spectrum at rest-frame 8 keV, the same energy as the peak in the lag–energy spectrum. To estimate the strength of the spectral feature, we fit it with a simple Gaussian function and find the equivalent width (that is, the area of the Gaussian above the continuum) to be $60 \pm 10$ eV. If this line corresponds to highly ionized iron K$\alpha$ emission at 6.97 keV, then the emission is blueshifted by a line-of-sight velocity of $\sim 0.15c$.

With our new timing results, we estimate the mass of the central black hole by comparing the lag amplitude and lag frequency of Swift J1644+57 to Seyfert galaxies with known reverberating broad iron lines. The time lag of the broad iron line is determined by the light travel time between the continuum-emitting region and the X-ray photosphere. The temporal frequency at which the iron K reverberation lag is found is determined by how fast the continuum is varying. Both the light travel time and the variability timeframe are expected to scale with the mass of the central black hole. These two correlations have been seen in Seyfert galaxies for which X-ray reverberation is detected[19,21]. In Fig. 3, we plot (as grey hexagons) the known correlations of the rest-frame temporal frequency of the iron K lag versus the black hole mass (Fig. 3a) and the rest-frame amplitude of the iron K lag versus black hole mass (Fig. 3b) for several Seyfert galaxies. The red solid lines in both panels show the best-fit line to these correlations using an orthogonal-distance regression fitting procedure to account for the error in both the $x$ and $y$ variables. For clarity, we do not plot the error bars on the mass in the figure, but they were included in the model fitting (see Extended Data Table 3 for the names and references of individual sources). The blue squares indicate the redshift-corrected temporal frequency (Fig. 3a) and lag amplitude (Fig. 3a) for our TDE source placed at the black hole mass suggested by the best-fit line. Both the lag amplitude and frequency correlations suggest a mass of a few million solar masses (as shown by the blue shaded region), confirming that this source achieves isotropic luminosities exceeding the Eddington limit by a factor of >10. The TDE fall-back rate for such a black hole mass is sufficient to produce a luminosity of $\sim 10^{46}$ erg s$^{-1}$ assuming a radiative efficiency of ~10%. See Methods for a discussion on the spread in this correlation, and the possible biases in mass that such a spread may imply.

The detection of iron K reverberation in this super-Eddington accretion event puts strong constraints on the geometry of the flow and the launching mechanism of the jet. It definitively shows that the optically thick accretion flow is being irradiated by a variable X-ray source. Although blazars sometimes show soft lags[22], there is no evidence of iron K reflection in either their lags or their time-integrated energy spectra. Therefore, the detection of iron K reverberation in Swift J1644+57 strongly disfavours a model in which the X-ray emission is produced by a highly relativistic (Lorentz factor >10), magnetically powered, blazar-like jet, as previously discussed[3]. Following recent theoretical developments in our understanding of super-Eddington accretion flows[23–26], we believe that we are looking down the polar funnel of a geometrically thick, radiation-supported flow (see Fig. 4 for a schematic). Therefore, the blueshift of the peak in the iron line is probably not due to disk rotation (requiring an edge-on disk), but rather due to the Doppler shift from outflowing material on the funnel wall that has been radiatively accelerated to $0.1c$–$0.5c$. In such a flow, a substantial part of the beaming is geometric (radiation preferentially escapes down the funnel) and the slow jet is probably launched by radiation pressure (see Methods and Extended Data Fig. 7 for an application of this simple model).

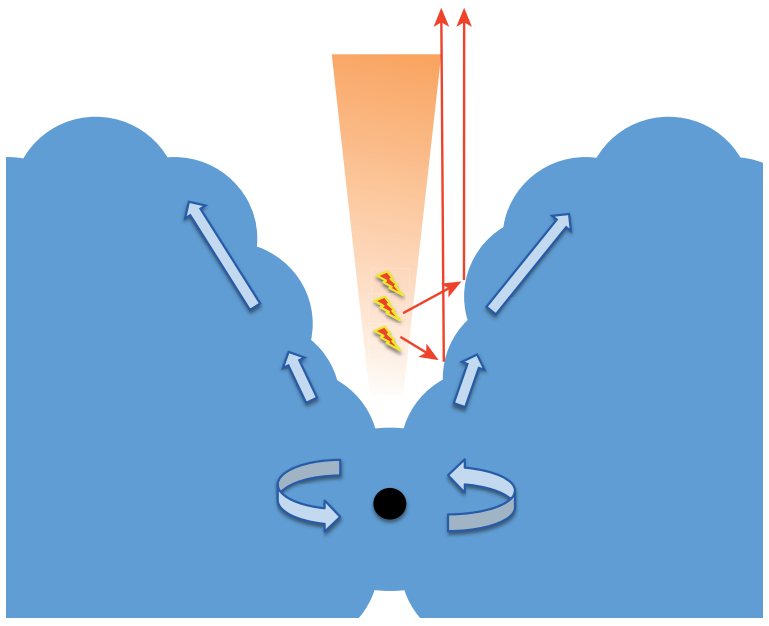

**Figure 4 | Schematic of reverberation off a super-Eddington accretion flow.** Swift J1644+57 is thought to be a super-Eddington accreting source, with a very thick, face-on accretion disk (blue) and a relativistic radio jet (orange). We suggest that the iron K lag originates from the light travel time between a flaring X-ray source and an irradiated outflowing funnel wall. In this schematic of the cross-section of the thick disk, the observer is looking down the funnel into the relativistic depths close to the central black hole. Variable X-ray continuum emission (lightning bolts) follows several light paths (two of which are shown as red arrows). The continuum emission reflects off the walls of the funnel and into the observer's line of sight. The strong blueshift of the iron lag is due to special relativistic Doppler shifts, and the asymmetric red wing is due to gravitational redshift from the strong gravity of the black hole. The blue arrows represent the dynamics in the disk: the accretion flow rotates around the central black hole and the walls of the funnel are outflowing at $\sim 0.1c$–$0.5c$.

Relativistic reverberation off an optically thick super-Eddington accretion flow tells us that much of the X-ray emission originates close to the event horizon. In principle, therefore, the X-ray emission can be used to measure the spin of the central black hole. However, our current models for measuring black hole spin from the broadening of the iron K emission assume a geometrically thin disk extending to the innermost stable circular orbit (ISCO), an assumption that is not likely to be relevant for this super-Eddington source, which has a very thick accretion disk that possibly extends beyond the ISCO to the marginally bound orbit[27,28]. Therefore, at this time we cannot claim an estimate of the black hole spin; but, with future developments in modelling such flows, relativistic reverberation from TDEs offers a potentially powerful technique for measuring spin—not only in the 10% of black holes that are persistently accreting, but also in the 90% of dormant black holes in the Universe.

**Online Content** Methods, along with any additional Extended Data display items and Source Data, are available in the online version of the paper; references unique to these sections appear only in the online paper.

**Acknowledgements** E.K. thanks A. Zoghbi, M. C. Miller, F. Tombesi, E. Miller and L. Denby for discussions. E.K. also thanks the Hubble Fellowship Program for support under grant number HST-HF2-51360.001-A from the Space Telescope Science Institute, which is operated by the Association of Universities for Research in Astronomy, Incorporated, under NASA contract NAS5-26555. J.M.M. acknowledges N. Schartel and XMM-Newton for executing target-of-opportunity observations of Swift J1644+57. C.R. acknowledges support from NASA under grant number NNX14AF86G. L.D. thanks J. McKinney for discussions. L.D. acknowledges support from NASA/NSF/TCAN (NNX14AB46G), NSF/XSEDE/TACC (TG- PHY120005) and NASA/Pleiades (SMD-14-5451). This work is based on observations made with XMM-Newton, a European Space Agency (ESA) science mission with instruments and contributions directly funded by ESA member states and the US (NASA) and the Suzaku satellite, a collaborative mission between the space agencies of Japan (JAXA) and the US (NASA).

**Author Contributions** E.K. led the XMM-Newton and Suzaku time lag analysis, simulations, interpretation of the results and manuscript preparation. J.M.M. performed the XMM-Newton spectral analysis and contributed to the interpretation of the results. C.R. developed the analytical toy model for reverberation in a super-Eddington flow and contributed to the interpretation of the results. L.D. had the idea of examining X-ray time lags in a TDE and contributed to the interpretation of the results.

**Additional Information** Reprints and permissions information is available at www.nature.com/reprints. The authors declare no competing financial interests. Readers are welcome to comment on the online version of the paper. Correspondence and requests for materials should be addressed to E.K. (ekara@astro.umd.edu).

## METHODS

**XMM-Newton data reduction.** XMM-Newton[29] observed Swift J1644+57 with the EPIC-pn camera[30] for 25 ks in large-window imaging mode with a medium optical block filer. We reduced the data using the XMM-Newton Science Analysis System (SAS version 14.0.0) and the newest calibration files. The events were filtered with conditions PATTERN ≤ 4 and FLAG = 0. See Extended Data Table 1 for a summary of the observations used in this analysis.

For the spectral analysis we used an annulus source region, excising the central pixels as the spectrum is clearly affected by pile-up. We excised the central 12 arcsec of the circular source region of 60 arcsec. We could not use the same source region for the detailed timing analysis because this source region spanned two chips that have slightly different readout times that affect the timing analysis. We therefore used a smaller circular source region of 35 arcsec contained on a single chip. For the timing analysis we did not need to excise the central pixels because pile-up does not affect the measurement of the lag. Pile-up is a random process and therefore does not contribute to the correlated variability between energy bands from which the lags are determined[31].

For both the spectral and timing analyses we used a circular background region with a radius of 70 arcsec. There were no strong background flares during the observation and the 0.3–10 keV background count rate (corrected for region size) never exceeded 0.4 counts per second, ~4% of the source count rate. The response matrices were produced using the rmfgen and arfgen tools in SAS.

We used the epiclccorr tool to produced background subtracted light curves with 10 s bins in several small energy bands in the range 0.3–10 keV. The 0.3–10 keV background-subtracted light curve is shown in Extended Data Fig. 2. The inset of Extended Data Fig. 2 shows a zoom-in of the light curves during the brightest, most variable part of the observation. We specifically examine the 4–6 keV light curve (red) and the 6–10 keV light curve (blue). By examining closely the light curves, one can see that on average the 4–6 keV band lags with respect to the 6–10 keV band (as shown in the lag–energy spectrum of Fig. 1). Although a lag is evident even in the time domain, the Fourier domain offers a clearer look at the lags on particular timescales.

We attempted to extract lags and spectra from an XMM-Newton observation taken shortly after the original one. Unfortunately, this observation was heavily affected by background flares, as noted previously[17]. Only 4 ks of the entire 21 ks exposure had a background level lower than the suggested maximum rate of 0.4 counts per second in the 10–12 keV background light curve. No iron K line or lag was detected, but given the poor quality of the data we cannot say whether this non-detection is intrinsic to the source.

**Fourier time lag analysis of the XMM-Newton data.** The Fourier time lag analysis follows the standard procedure[12,32]. In short, for every energy bin shown in the lag–energy spectrum (Fig. 1), the time lag is computed between the light curve in that particular bin of interest and the large reference-band light curve from 0.3–10 keV. We remove the bin of interest from the reference band so we do not have correlated errors. As the lag is measured in each bin of interest relative to essentially the same reference band, the lag–energy spectrum displays the relative lag between different energy bins.

To compute the lag we take the Fourier transform of the light curve and the reference-band light curve and compute the cross spectrum by multiplying one band and the complex conjugate of the other band. The argument of the cross spectrum is the phase difference between the two bands. For the lag–energy spectrum, we focus on the average phase lag over a particular frequency range. We convert the phase lag into a time lag by dividing by the midpoint frequency. In practice, the lag–energy spectrum is read from bottom to top—the bins with the smallest lag respond first, whereas the bins with larger lags are delayed with respect to the other bins.

We choose the frequency range in the usual way, on the basis of the lag–frequency spectrum between rest-frame 0.3–1 keV and 1–4 keV (Extended Data Fig. 1). These bands are chosen because the 0.3–1 keV band is typically the band dominated by the soft excess and the 1–4 keV band is typically the band most dominated by the continuum[9,19]. We find a negative lag (that is, the soft band lagging the hard band) from $(2–10) \times 10^{-4}$ Hz and therefore our lag–energy analysis is focused on this particular frequency range.

The power of the cross-spectral Fourier technique is that we pick out variability features that are correlated between energy bands. This, in principle, allows us to pick out components that may be lost in the time-averaged spectrum.

**Suzaku data reduction.** We process the unfiltered event files for each of the XIS CCDs. The source and background regions were ~3.5 arcmin in radius. XSELECT was used to extract the spectral products, and medium-resolution responses were generated using XISRESP. The front- and back-illuminated light curves were combined for the time lag analysis. We extracted light curves between 1–8 keV in 1 keV energy bins and 128 s time bins. We include the 1.8–2.2 keV band where there are known calibration uncertainties because they do not affect the time lag analysis. The signal was too low in the 8–10 keV bins, and the maximum-likelihood analysis could not converge. We therefore exclude these highest-energy bins from the analysis. The total Suzaku exposure is 52 ks.

The Suzaku spectrum does not show a clear iron K emission line. We find an upper limit of 7 eV on an iron line of equivalent width at the same energy as the XMM-Newton spectrum. We closely examined the spectrum above 3 keV, ignoring the Au edge at ~2.3 keV, which resulted in additional residuals to the fit. We find that the 3–10 keV spectrum shows slight hardening above ~6 keV. This hardening is not likely to be due to pile-up because the observation was taken in the one-quarter window mode and the observed count rate is well below the nominal pile-up limit of 100 incident counts per second for the one-quarter window mode. Furthermore, excising the central 20 arcsec of the source region does not remove the hardening. We checked the Suzaku Hard X-ray Detector (HXD) spectrum and found that the excess hardening was also present above 20 keV. Extended Data Fig. 3 shows the ratio of the Suzaku X-ray Imaging Spectrometer (XIS) and HXD data to an absorbed power-law fit to the 3–6 keV range (this includes a normalization cross-calibration constant of 1.16 between front-illuminated XIS and HXD, as suggested in the Suzaku ABC guide). This ratio plot displays the hard tail above ~6 keV and the excess above 20 keV. The excess hardening and curvature above 20 keV are reminiscent of the Compton hump. The presence of a Compton hump without the accompanying detection of an iron line is puzzling, but could suggest the presence of an iron line that is so severely broadened that it is rendered undetectable. This in turn would suggest that the reflection features were produced much closer to the black hole (and that our line of sight extended down to this region) during the Suzaku observation compared with the later XMM-Newton observation.

**Maximum likelihood analysis of Suzaku data.** As Suzaku is in a low-Earth orbit there are gaps in the observations approximately every 3,000 s. The timescales relevant to this system are longer than 3,000 s and therefore we cannot rely on the Fourier method. Instead we chose the maximum-likelihood method[33], which is used to analyse the lags in NuSTAR observations[34,35]; for further details of the method and Monte Carlo simulations that compare it with the Fourier method see the indicated references.

The Suzaku lag–energy spectrum is shown in Extended Data Fig. 4. The lag increases up to the same at the same energy as in the XMM-Newton observation (8 keV rest frame) and then drops again at higher energies. We note that the iron K lag amplitude that we measure is roughly the same size as the time bin of the light curve. The lag amplitude that we measure is an average time lag, and therefore the observed lag amplitude is not directly set by the bin size. The only clear difference between the Suzaku lags and those found in XMM-Newton are from 1–2 keV. The 1–2 keV bin in the Suzaku data are 110 ± 50 s less than the XMM-Newton lags. We hesitate to over-interpret one bin, but note that differences in the soft excess are also seen between Seyfert galaxies that show similar iron K lag features[19].

This analysis shows that we detect iron K reverberation in a TDE in two observations taken at different times, made with different observatories and analysed using different techniques.

**Significance of the lag.** In this section we present several statistical tests to confirm the detection. We start by testing whether the observed lags are due to random fluctuations in the data. Next we test whether the observed lags are consistent with the behaviour of the iron K lags seen in other Seyfert galaxies. Once we determine the significance of the detection of the iron K lag, we test whether this line profile is broad and asymmetric, as these features have important implications for the interpretation of the iron K emission originating from close to the central black hole.

We start by testing the detection of the lags using Monte Carlo simulations. We simulate light curves using the method described in ref. 36. We obtain our underlying power spectral density model (PSD) by fitting the 0.3–10 keV observed PSD with a broken power law (plus a constant for the white noise). With this relatively short observation, we do not observe a break in the PSD and so we assume a break frequency of $10^{-4}$ Hz to a low-frequency power law with index 1 (consistent with previous results[37] on the low-frequency variability of Swift J1644+57). Above $10^{-4}$ Hz we find a high-frequency power-law index of 2.1.

We simulate 10,000 light curves in each energy bin, scaled to the observed variance, count rate and exposure for that bin. We compute the time lag between each bin and the large reference band and assume zero intrinsic lag. Extended Data Fig. 5 shows the 1$\sigma$ distribution of the lags in each energy bin overlaid with the observed XMM-Newton lag–energy spectrum to check the probability that the observed 0.3–10 keV lag–energy spectrum could be due to random fluctuations. Summing $\chi^2 = (x-\mu)2/\sigma^2$ for each independent energy bin (where $x$ is the observed lag, $\mu$ is the mean of our zero-lag null hypothesis and $\sigma$ is the standard deviation based on the Monte Carlo simulations), we obtain $\chi^2$/d.o.f. = 33/17.

Our simulations therefore rule out a zero-lag null hypothesis with >99% confidence. This statistical test does not, however, account for the fact that the lags are coherently distributed above and below the zero-lag null hypothesis and not simply randomly distributed. The distribution of lags is Gaussian for all energy bins.

Next we test whether the 3–10 keV lags that we observe are associated with iron K emission lagging behind the continuum. Most sources with iron K lags show the smallest lag at the red wing of the iron line, a gradual increase in the lags up to the line centroid and a sharp drop above the line centroid back to small lags[12]. Phenomenologically, this behaviour is well described by a power-law continuum plus an asymmetric and broad emission line peaking at the line centroid. We use the diskline model as a phenomenological model to test for the presence of an iron K lag similar to those seen in Seyfert galaxies, however, we emphasize that the diskline (which describes the line broadening from a rotating, inclined, thin accretion disk) is formally not an appropriate model to describe emission from a face-on thick accretion flow. For the purposes of this simple statistical test for the presence of an iron K line in the lag–energy spectrum, the diskline model is simply employed in its role as a 'generic' skewed line profile and is sufficient.

To test for the presence of the iron K lag in Swift J1644+57, we compare two nested models in the 3–10 keV band: one of only a constant fitted to the 3–10 keV lag–energy spectrum and other containing an additional diskline model. For the diskline model we leave the line energy and inclination free (to allow some freedom in the width of the line profile) and freeze the inner and outer radii to $6r_g$ and $200r_g$, respectively (where $r_g = GM/c^2$ is the gravitational radius), with a power-law disk emissivity. We use an $F$-test to compare two nested models. Because the normalization of the lag can be negative, we avoid the problem of fitting too close to the parameter space boundary[38]. The resulting $F$-statistic value is $\Delta\chi^2/\chi^2 = 23.8$ and therefore the constant plus diskline model is preferred with >99.9% confidence. If we assume that the iron K lag found in Suzaku data has the same diskline energy and inclination as the XMM-Newton lag, then the $F$ test indicates that the constant plus diskline model is preferred to >99.8% confidence. The same test with a Gaussian rather than a diskline model results in the Gaussian being preferred at >99.2% confidence for the XMM-Newton data and 98.8% confidence for the Suzaku data. This suggests that an asymmetric diskline model is preferred over the Gaussian, though this needs to be tested rigorously (see below).

Now that we have tested for the presence of an iron K lag, similar to those seen in Seyfert galaxies, we test whether the iron line is broad and whether the iron line is asymmetric. For the first test, we simply use the function steppar in XSPEC, which steps through the parameter space to calculate $\Delta\chi^2$ at each step. We find that the Gaussian is inconsistent with a narrow line at >99% confidence. Next we compare the best-fit broad Gaussian model with the asymmetrically broadened diskline model. Although the diskline model provides a lower $\chi^2$ value for the same number of degrees of freedom, both models result in formally acceptable fits (that is, result in a reduced $\chi^2$ less than 1). See a comparison of the Gaussian and diskline models in the right panels in Extended Data Fig. 5. We follow a Bayesian formalism[43] to quantify the improvement of the asymmetric diskline. To compare the broad Gaussian fit with the asymmetric diskline model, we compute the ratio of the evidences—the so-called Bayes' factor[38]. The Bayes' factor is the ratio of the posterior probabilities of our two models given the observed data, which we determine by stepping through the free parameters of each model and computing $\chi^2$ at each point. We convert $\chi^2$ into a probability for each step and sum the probabilities over the domain (in effect, integrating the probability over our defined parameter space). The free parameters and their domains for each model are shown in Extended Data Table 2. For simplicity, we freeze the normalization of the power law for both models to be the average lag from 3–4 keV and 6–10 keV. We find the Bayes' factor (the ratio of the diskline-integrated probability to the Gaussian probability) to be 55. This corresponds to the asymmetric diskline being favoured over the Gaussian at >98% confidence.

**Comments on the reverberation lag–mass relation.** In Fig. 3, we compared the frequency and amplitude of the iron K lag with the black hole mass for several Seyfert galaxies with known reverberation lags. Those sources are shown in Extended Data Table 3. We use all published iron K lag results, excluding the iron K lag in MS 22549−3712, which is associated with a quasiperiodic oscillation and may not have the same variability mechanism as the other sources[20].

Where possible, we use optical reverberation mass estimates with scale factor $\langle f \rangle = 4.3$ (ref. 39). These were obtained from the AGN (active galactic nuclei) Black Hole Mass Database (http://astro.gsu.edu/AGNmass). Most masses were calculated using the relation between the broad line region (BLR) radius and the luminosity, $R_{BLR} \propto L^{\alpha}$ (refs 40–42), with $\alpha = 0.5$. For sources that do not have an optical reverberation mass estimate, we assumed an error of 0.5 dex.

Some of the spread in the correlation between the lag amplitude and black hole mass in Seyfert galaxies is probably due to differences in the accretion flow geometry and orientation. To this effect, if the geometry of the tidal disruption is very different to that of Seyfert galaxies (as we expect), then Swift J1644+57 may not lie on this correlation. Therefore, some caution should be exercised in using the lag amplitude as a definitive indicator of the mass. We do note however, that although the amplitude of the lag is dependent on the geometry of the flow (that is, dilution effects, different light travel times, the effect of inclination), the frequency at which the lag is found is less dependent on geometry and is therefore likely to be a more trustworthy indicator.

In Extended Data Fig. 6 we compare the reverberation lags of three rapidly variable Seyfert galaxies and the TDE Swift J1644+57. The sources are 1H 0707−405, IRAS 13224−3809 and Swift J2127.4+5654. To compare the shapes of the lag–energy spectra of different mass objects, we have scaled the lags so that they have the same iron K lag amplitude. There are clearly similarities in the overall spectra (for example, the lag of the soft excess, broad iron K line and Compton hump with respect to the continuum), although there are differences in the shapes of the line profiles. Swift J644+57 has a slightly narrower iron K line profile than 1H 0707−495 and IRAS 13224−3809. The energy of the line centroid in Swift J1644+57 is also slightly higher than the line centroid in Swift J2127.4+5654. We discuss possible physical reasons for this in the next sections.

**Theoretical framework of super-Eddington accretion in TDEs.** During a TDE the black hole goes from being dormant to radiating at or above the Eddington luminosity within around one month, and then decays in around a year to sub-Eddington levels[13]. The dynamical mass fallback rate of the debris is super-Eddington when the black hole mass is less than 50 million solar masses[13]. If the circularization of the debris is efficient enough[44–46], then the accretion on the black hole is also super-Eddington.

The small super-Eddington accretion disk in a TDE has not been well studied, but very recently numerical simulations have been employed to understand super-Eddington disks in persistent sources such as AGN or X-ray binaries[23–25]. Although the overall radiative efficiency and luminosity are still debated, in all simulations the disk structure is geometrically thick and optically thick, deviating from the standard thin disk picture[47]. A strong optically thick outflow is also generated and radiation can only leak through a narrow funnel along the polar direction. Close to the black hole, simulations show that a jet can help to carve out the inner accretion flow, thus exposing the X-ray emitting region of the disk[25]. In the final section, we present a toy model of the X-ray emission based on the geometry suggested by these numerical simulations.

**Iron line reverberation from the outflowing eye-wall of a super-Eddington accretion flow.** XMM-Newton shows that the iron line peaks at 7.5–8 keV in both the direct energy spectrum and in the lag–energy spectrum. Even in the case of hydrogen-like iron (with an iron-K$\alpha$ line energy of 6.97 keV) the observed energies necessitate a significant blueshift. In the case of sub-Eddington AGN that possess geometrically thin accretion disks such blueshifts would suggest a high-inclination accretion disk. In the case of Swift J1644+57, however, multiple lines of evidence[4,5] point to a low-inclination (pole-on) geometry. Given the robust theoretical expectation (discussed in the previous section) that the accretion disk of such a super-Eddington source should be very geometrically thick, we suggest that the blueshifts seen in the iron line correspond to (radiation-driven) outflows in the funnel, or 'eye wall', of a geometrically thick accretion flow. In addition to the blue wing, the strong red wing of the iron line necessitates gravitational redshift (neglecting the small effect of multiple scattering events that can Compton downscatter some iron K photons to lower energies). This therefore suggests that the base of the geometrically thick funnel reaches deep into the potential well of the black hole, within 10 gravitational radii.

As proof of principle we construct a simple toy model of iron line reverberation in a funnel geometry (see Fig. 4). The funnel is assumed to be conical with a half-opening angle of $\theta$, and we consider X-ray line reverberation driven by some variable X-ray source on-axis in the funnel at distance $R_X$ from the black hole. We suppose that line emission can be excited along the walls of the funnel from radii $R_{min}$ (within which we assume the accretion flow to be completely optically thick) to $R_{max}$ (beyond which the disk starts to flatten off and cannot see the X-ray source). In this toy model, we neglect the rotation of the flow and simply prescribe a radial outflow velocity profile for the wall material:

$$v(R) = v_0 + (v_t - v_0)\frac{R - R_{min}}{R + R_{acc}}$$

where $v_0$ is an initial velocity, $v_t$ is a terminal velocity, and $R_{acc}$ is the characteristic radius over which the acceleration proceeds. We then calculate the line profile in both the flux–energy and lag–energy spectra, assuming that we are viewing the funnel exactly pole-on. We include the effects of the relativistic Doppler shifts, relativistic beaming and gravitational redshift on both the excitation of the line emission by the source and the observed energy/lags. We do not, however, account for general relativistic light bending.

Example results for a range of terminal velocities are shown in Extended Data Fig. 7 for the following illustrative parameters: $\theta = \pi/4$, $R_{\rm min} = 6r_{\rm g}$, $R_{\rm max} = 200r_{\rm g}$, $R_{\rm X} = 30r_{\rm g}$, $v_0 = 0$. To calculate time-lags in physical units, we also need to assume a black hole mass: we take $M = 2 \times 10^6 M_\odot$. Strongly blueshifted line peaks can readily be obtained—with this parameter set, we find that a terminal velocity of $v_{\rm t} \approx 0.5c$ is required to have the line peak at 7.8 keV. We also obtain lag–energy spectra comparable to those seen in the data, peaking at 7.8 keV with a tail to lower energies. We obtain qualitatively similar results if we have a narrower opening angle (that is, $\theta = \pi/6$) and slower terminal velocity ($v_{\rm t} \approx 0.3c$). Because of this degeneracy, we do not make a strong claim of the funnel opening angle.

An important technical point to note is that the time-lags derived in a reverberation analysis are smaller than the actual time-lags (that is, the light travel time of the extra path-length traversed by the reflected emission) due to dilution of the line-band by the coherent component of the direct continuum emission. In the toy model results shown here, we derive the observed (diluted) lags by multiplying the actual lags by the fraction that the line emission contributes at each energy, normalizing so that the peak of the line is 30% of its local continuum (motivated by the observed total energy spectrum, Fig. 2). It remains a distinct possibility that some fraction of the observed X-ray continuum is incoherent, meaning that we have overcorrected for dilution and hence underpredicted the dilution-corrected lags. A full treatment of dilution in any theoretical model requires a consideration of the full reflection spectrum and a Fourier-domain treatment that parallels that applied to the real data.

**Code availability.** The model fitting of spectra and lag–energy spectra was completed with XSPEC, which is available at the HEASARC website (http://heasarc.gsfc.nasa.gov). The code used for the time lag analysis and Monte Carlo simulations is not currently available. All figures were made in Veusz (http://home.gna.org/veusz).

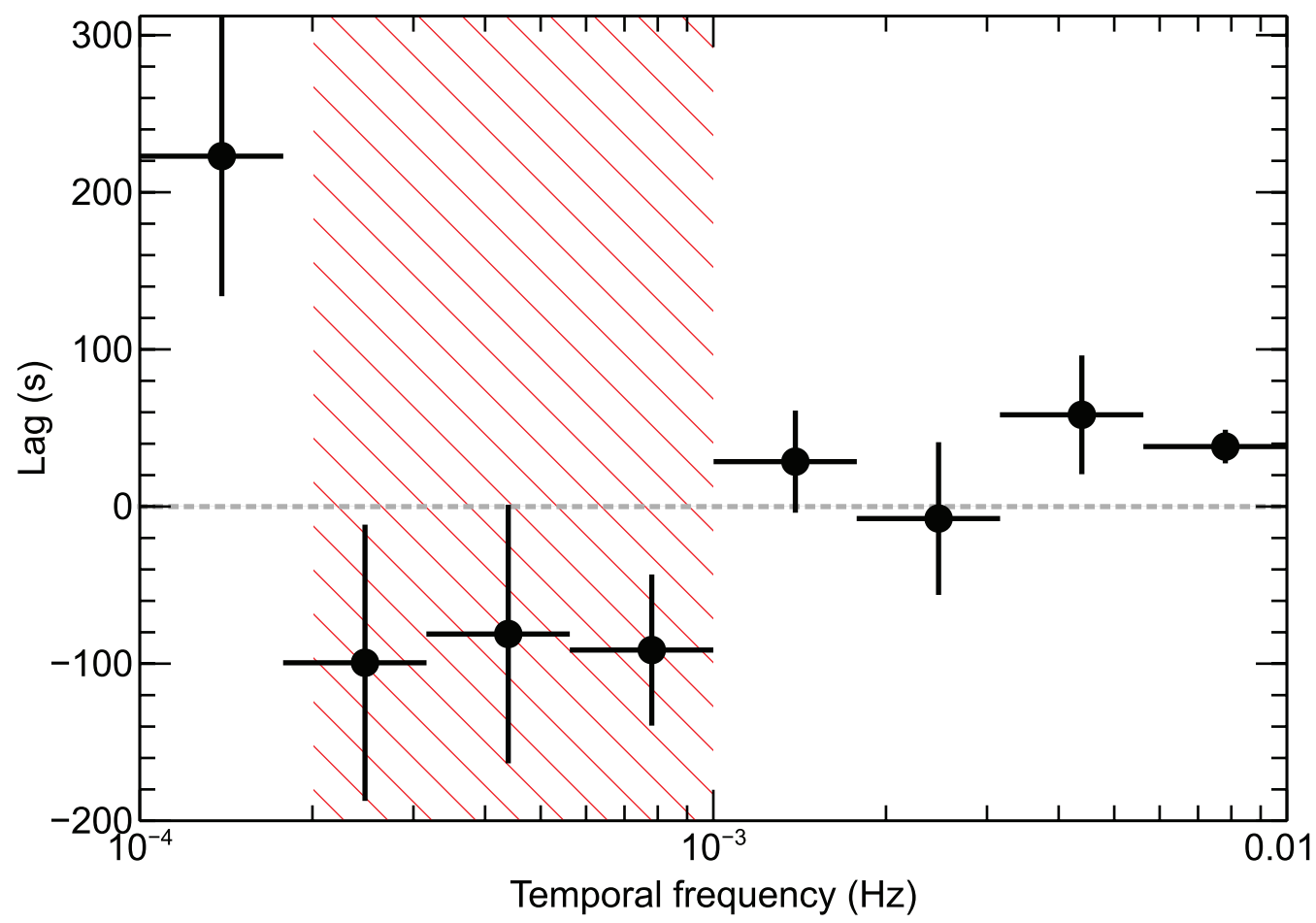

**Extended Data Figure 1 | XMM-Newton lag–frequency spectrum.** The lag–frequency spectrum between rest frame 0.3–1 keV and 1–4 keV (the usual soft and hard bands chosen for reverberation studies[9,21]). The lag–frequency spectrum shows a negative lag (that is, the soft band lagging behind the hard band) from $(2–10) \times 10^{-4}$ Hz (highlighted by the red hashed region). Because the lag–frequency spectrum shows a soft lag at these frequencies, we examined further the lag–energy spectrum at this particular frequency range (see Fig. 1). Error bars are at the 1$\sigma$ level.

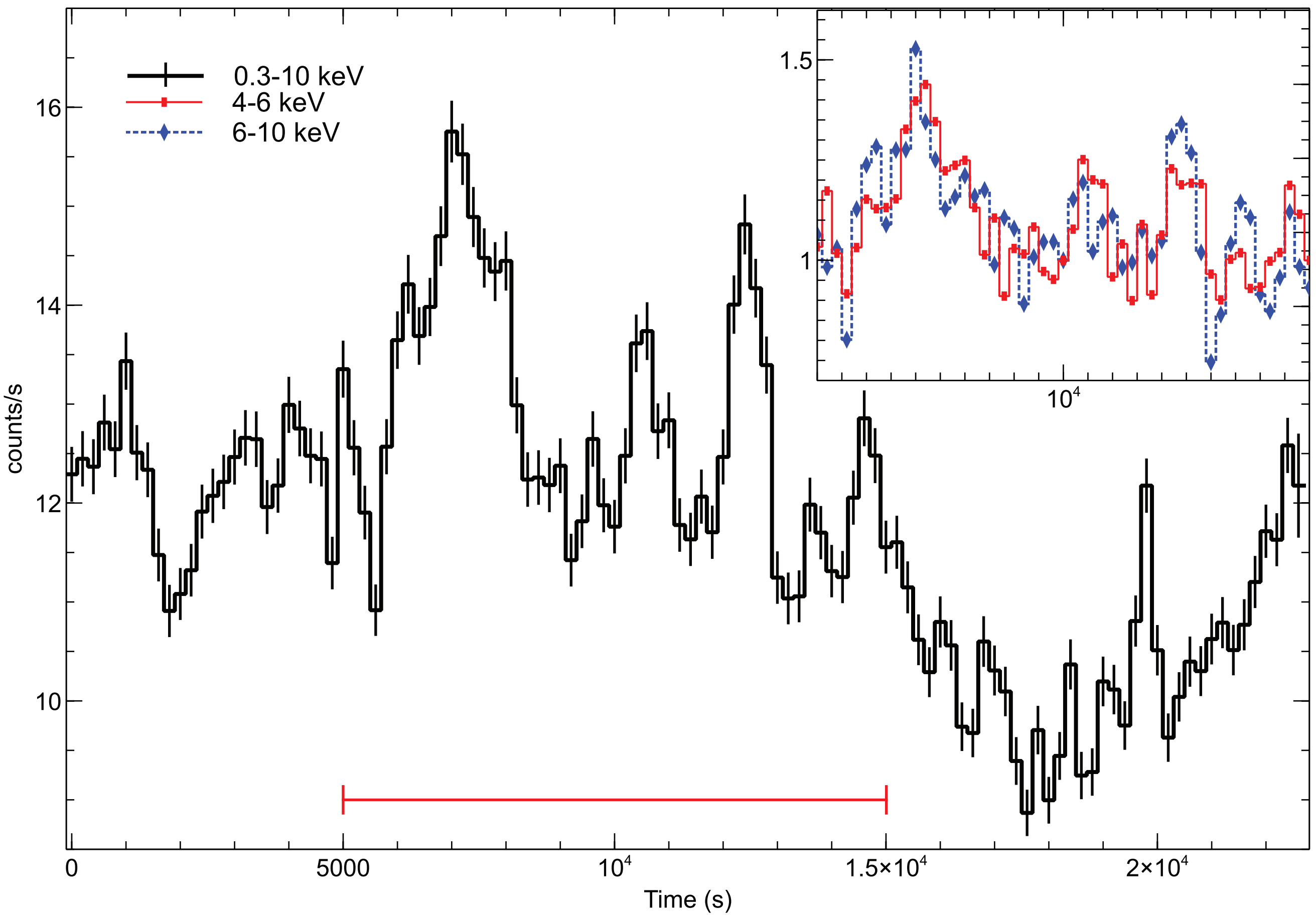

**Extended Data Figure 2 | XMM-Newton light curve of Swift J1644+57.** The observed 0.3–10 keV light curve in 200 s bins showing the rapid variability in this 25 ks observation. Note that the *y* axis does not extend to zero count rate. Error bars are at the 1$\sigma$ level. Inset, the 4–6 keV (∼5.5–8 keV rest frame) light curve (red) and the 6–10 keV (∼8–13 keV rest frame) light curve (blue) zoomed in on the most variable part of the light curve from 5,000 s to 15,000 s (indicated by the red line in the main figure). The *y* axis is in units of count rate divided by the mean of the entire light curve. This illustrates (in the time domain) that on average the 6–10 keV variability leads the 4–6 keV variability. We emphasize that the time delay seen in these coarsely binned light curves places an upper limit on the actual amplitude of the lag. The detailed Fourier analysis (Fig. 1) allows us to obtain a better estimate of the average time lag.

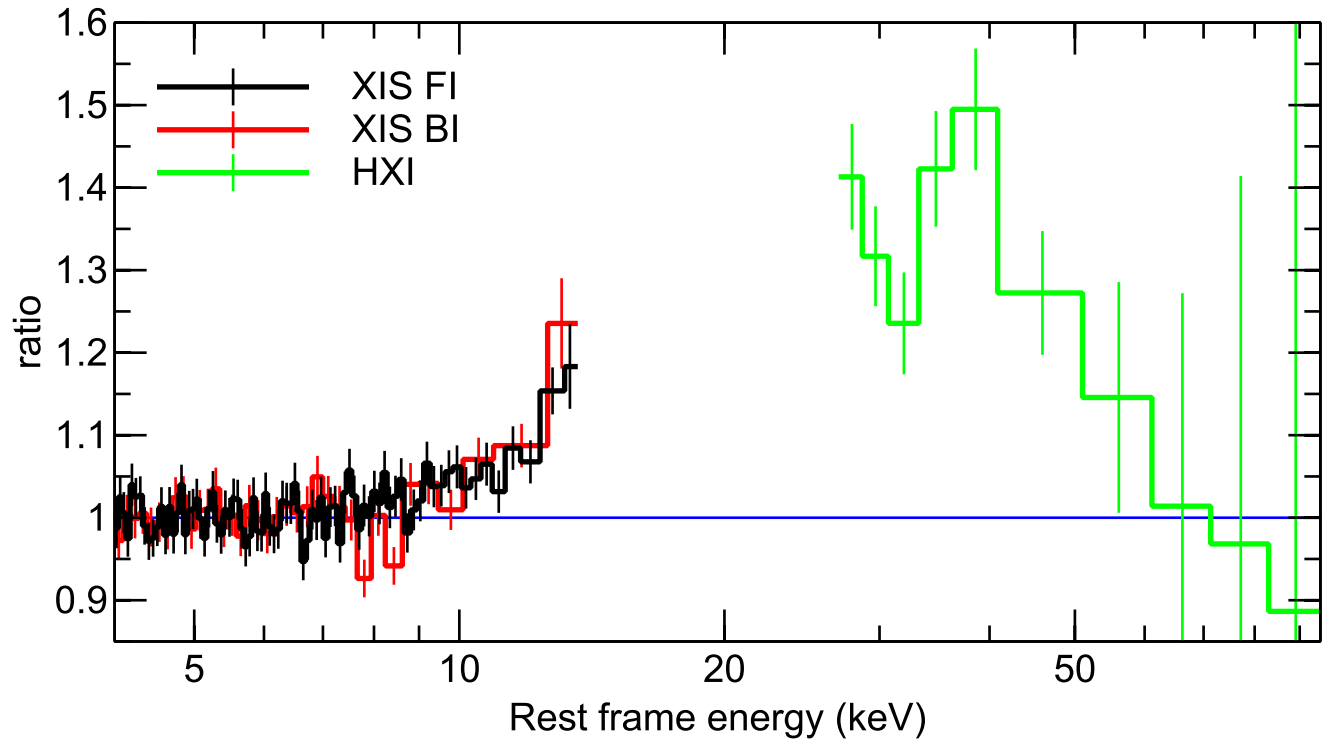

**Extended Data Figure 3 | Suzaku XIS and HXD spectra.** A ratio plot of the time-integrated energy spectra of the front-illuminated (black) and back-illuminated (red) XIS detectors and the HXD (green) to an absorbed power-law fit from 3–6 keV. The blue line at a ratio of one indicates where the data would be fully described by the absorbed power-law model. Error bars are at the $1\sigma$ level. This fit exhibits the potential hardening in the spectrum above $\sim$6 keV that appears to continue up to $\sim$30 keV, beyond which the spectrum turns over. There is no obvious iron line in the XIS spectra.

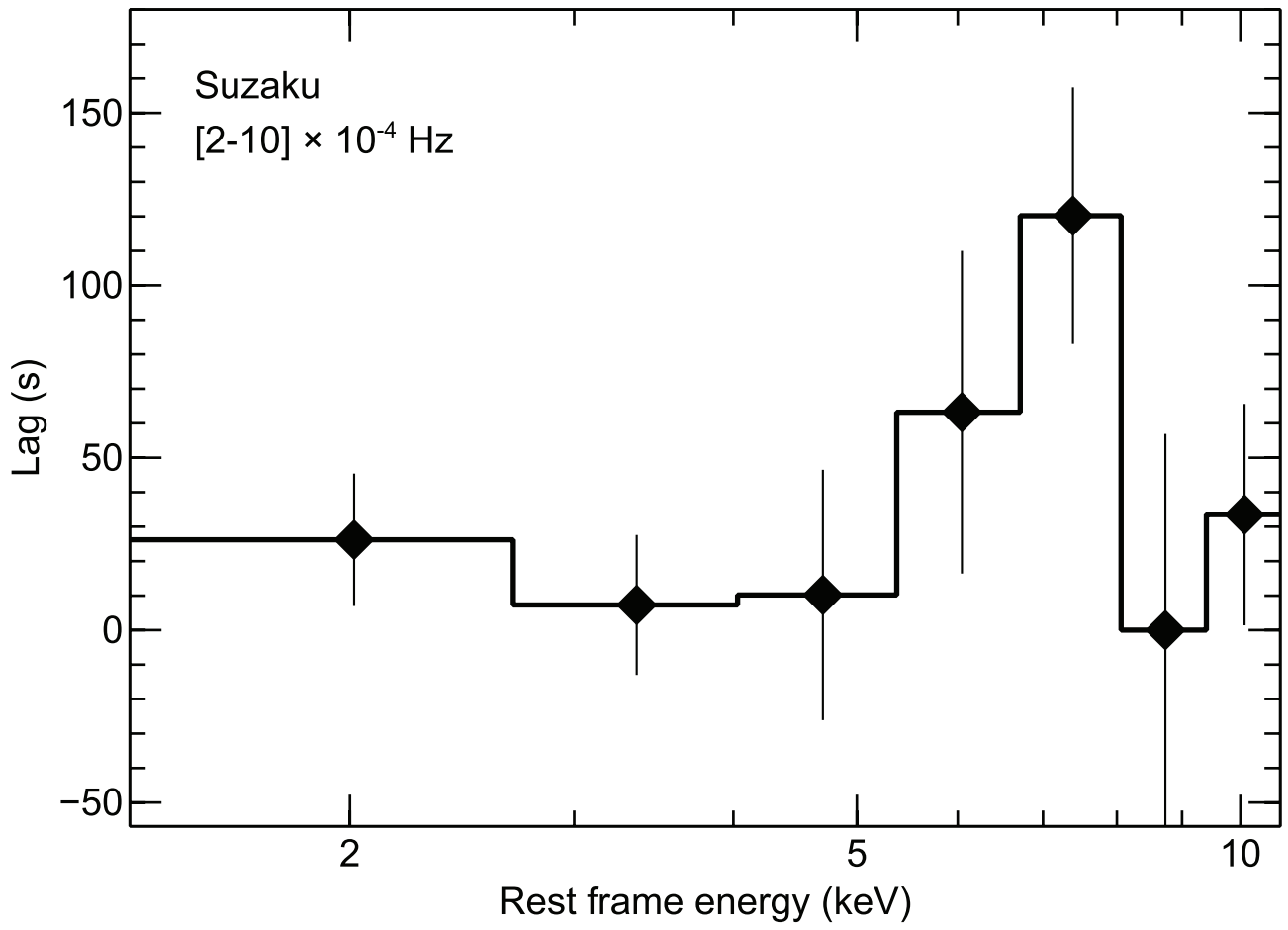

**Extended Data Figure 4 | Suzaku lag–energy spectrum.** The lag–energy spectrum of the Suzaku observation taken 10 days before the XMM-Newton observation. An iron K lag at the same frequency and same energy as the XMM-Newton iron K lag is detected in the data. This confirms the presence of an iron K lag with an amplitude of $\sim$120 s in two separate observations, taken with different instruments and analysed using different techniques. Error bars are at the 1$\sigma$ level.

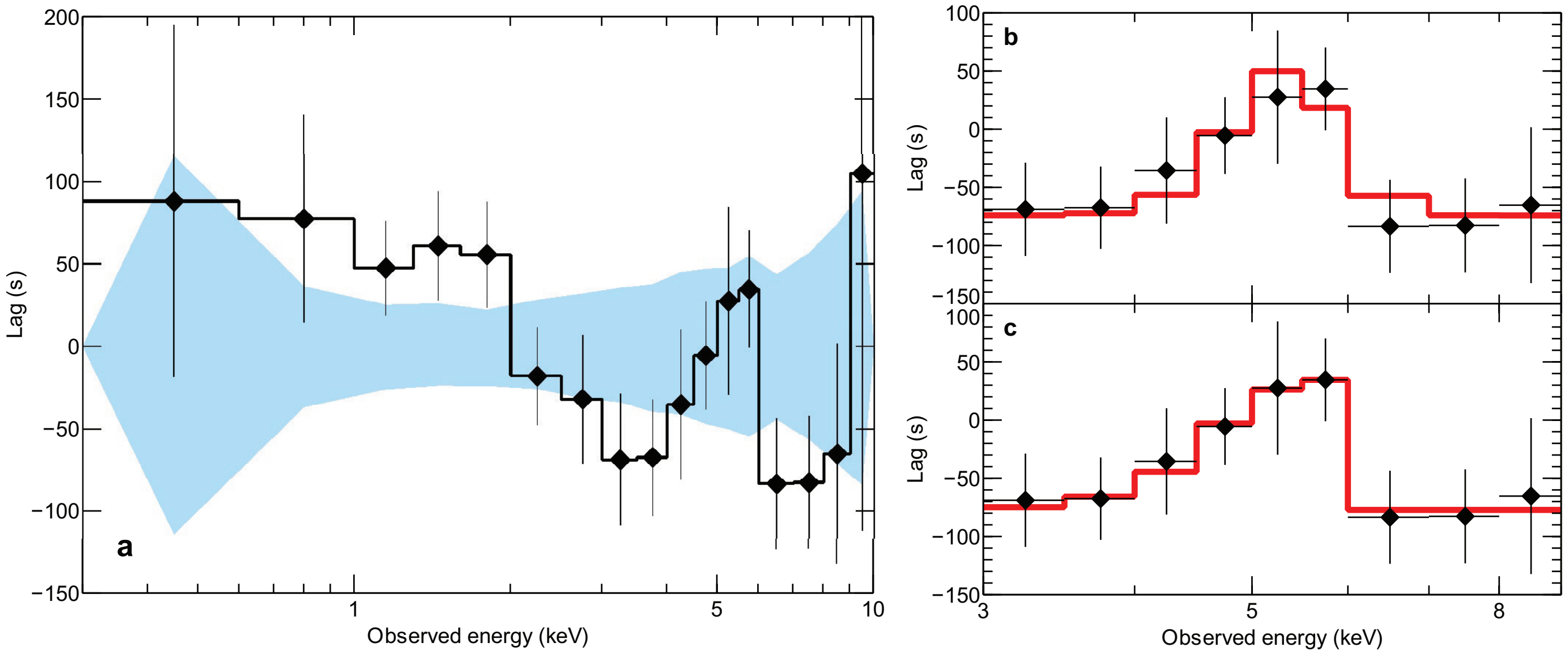

**Extended Data Figure 5 | Statistical significance of the XMM-Newton lag–energy spectrum. a**, The observed lag–energy spectrum (black points; same as Fig. 1 except the zero-point lag has not been shifted) compared with the 1$\sigma$ distribution of 10,000 simulated Monte Carlo light curve pairs in each energy bin with zero lag (blue shaded region). This plot shows that the observed 0.3–10 keV lag energy is inconsistent with zero lag at >99% confidence. See Methods for details of the simulations. **b**, **c**, The best Gaussian (**b**) and asymmetric diskline (**c**) model fits (red lines) to the 3–10 keV lag–energy spectrum. We compute the Bayes' factor between the models and find that the diskline is preferred at >98% confidence.

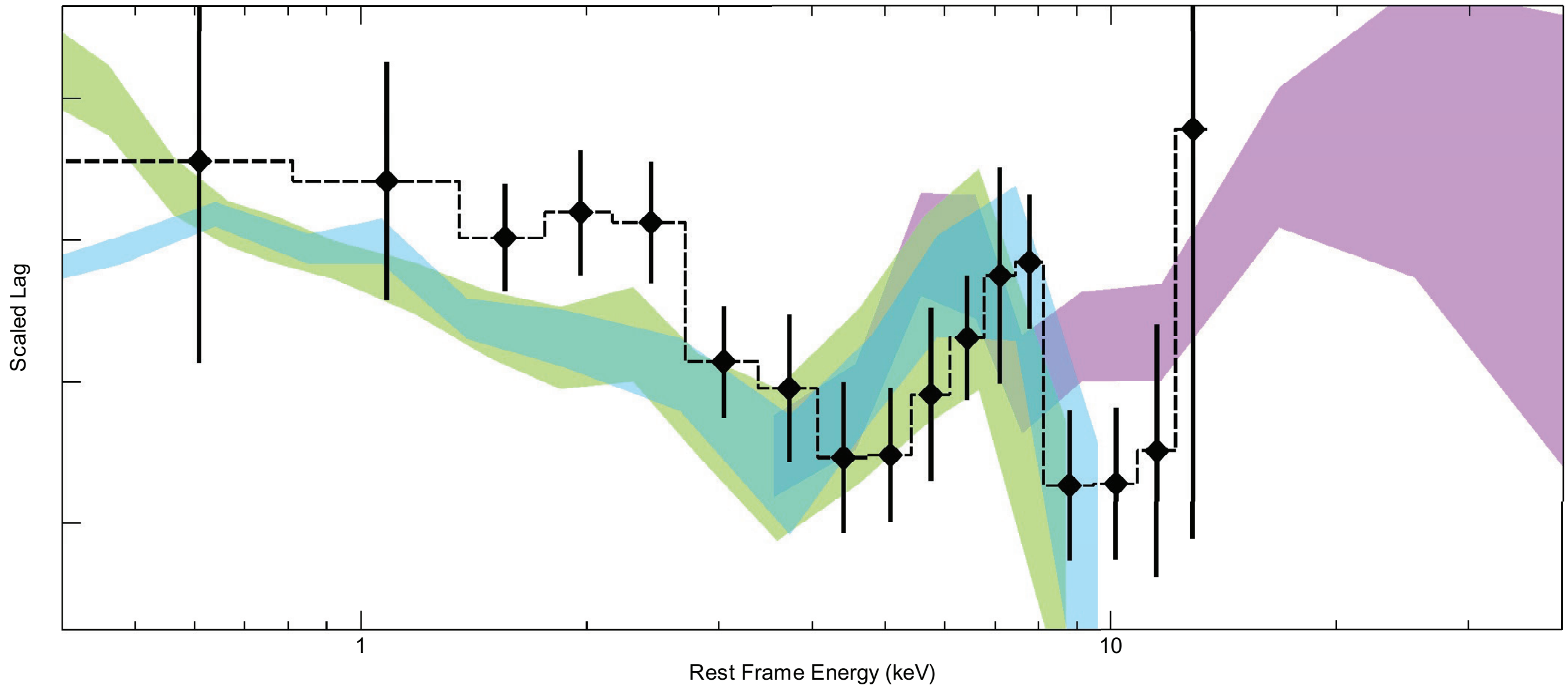

**Extended Data Figure 6 | Swift J1644+57 lag–energy spectrum compared with Seyfert galaxies.** Overlay of the reverberation signatures in nearby variable Seyfert galaxies compared with Swift J1644+57 (black diamonds). We show IRAS 13224−3809 in green, 1H 0707−495 in blue and Swift J2127.4+5654 in purple. The *y* axis shows arbitrary time lag units that have been scaled so that all of the sources have the same iron K lag amplitude. We show the scaled lag because all of the sources have different black hole masses (see Fig. 3). There is similarity in the broadband shape, although the iron K line profile of Swift J1644+57 appears to be slightly narrower and peaks at higher energies.

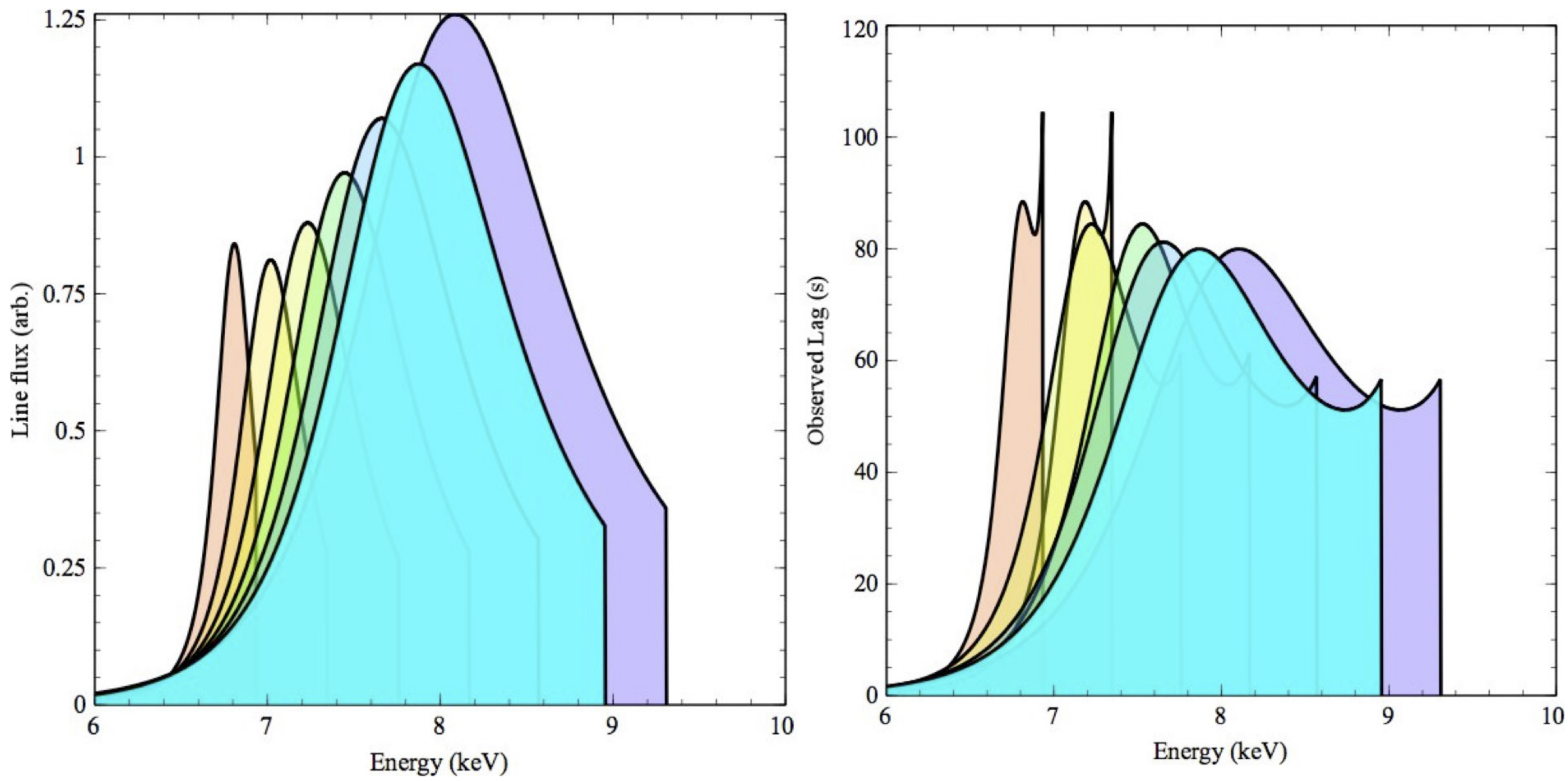

**Extended Data Figure 7 | Flux–energy and lag–energy spectra of reflection in super-Eddington flow.** Flux–energy (left) and lag–energy (right) spectra derived from our toy model for iron XXVI (rest-frame 6.97 keV) Kα line reverberation from an outflowing funnel wall. The results for terminal velocities of 0, 0.1*c*, 0.2*c*, 0.3*c*, 0.4*c*, 0.5*c* and 0.6*c* are shown (from left to right in peak energies).

**Extended Data Table 1 | Observations used in this analysis**

| Instument | Date (YYYY-MM-DD) | OBSID | Exposure (s) |
|---|---|---|---|
| *XMM-Newton* | 2011-04-16 | 0678380101 | 25000 |
| *Suzaku* | 2011-04-06 | 906001010 | 52000 |

A summary of the data used in this work, including the name of the telescope, the date of the observation, the OBSID unique identifier and the observation exposure length.

**Extended Data Table 2 | Results of the fits to the lag–energy spectrum**

| Gaussian Model | | | Diskline Model | | |
|---|---|---|---|---|---|
| **Parameter** | **Best-fit** | **Domain** | **Parameter** | **Best-fit** | **Domain** |
| Line energy (keV) | 5.3 | [3, 10] | Line energy (keV) | 5.3 | [3, 10] |
| Width (keV) | 0.67 | [0, 3] | Inclination (deg) | 42 | [0, 90] |
| Normalization (s) | 188 | [0, 500] | Normalization (s) | 167 | [0, 500] |
| **Bayes' Factor ($P_{\text{diskline}}/P_{\text{Gaussian}}$):** 55 | | | | | |
| **Probability:** $> 98\%$ | | | | | |

The free parameters in the Gaussian and diskline models. We show the best-fit parameters, the domain over which we step through the parameters, the corresponding Bayes' factor and the probability. Note that for fitting in XSPEC we have shifted the lag amplitudes up by a constant value of 1,000 s so that none of the normalizations would be negative.

**Extended Data Table 3 | Lags and masses of Seyfert galaxies with reverberation detections**

| Name | Lag Amplitude (s) | Lag Frequency ($10^{-4}$ Hz) | Mass ($\log(M/M_{\odot})$) |
|---|---|---|---|
| **NGC 4151** [10,48] | 880±360 | 0.1–0.5 | 7.55±0.05 |
| **1H 0707-495** [49,50] | 47±16 | 10–30 | 6.4±0.5 |
| **IRAS 13224-3809** [51,52] | 299±135 | 1.4–2.8 | 6.6±0.5 |
| **MCG-5-23-16** [18,52] | 1037±455 | 0.1–10 | 7.9±0.5 |
| **NGC 7314** [18,52] | 77±30 | 0.1–4 | 6.9±0.5 |
| **Ark 564** [19,53] | 92±65 | 3.2–5.2 | 6.4±0.5 |
| **Mrk 335** [19,54] | 193±98 | 1.9–6.7 | 7.23±0.04 |
| **PG 1244+026** [55,56] | 762±306 | 0.9–3.6 | 7.3±0.5 |
| **Swift J2127.4+5654** [57,58] | 407±127 | 0.4–4.5 | 7.2±0.5 |
| **NGC 1365** [35,59] | 500±120 | 0.9–1.9 | 7.7±0.5 |
| **NGC 4051** [60,61] | 90±30 | 3.5–40 | 6.1±0.1 |

The observed iron K lag amplitude (measured between 3–4 keV and 5–7 keV) with $1\sigma$ errors, observed iron K lag frequency range and black hole mass with $1\sigma$ errors are shown for all Seyfert galaxies with known iron K reverberation detections. The two references listed in brackets after each object correspond to the reference[10,18,19,35,49,51,55,57,60] for the iron K lag amplitude and frequency, and the reference[48,50,52–54,56,58,59,61] for the black hole mass. These data correspond to the plot shown in Fig. 3 (although the error on the mass has not been plotted, for clarity).